\documentclass[12pt]{article}
\input{psfig}
\newcommand{\beq}{\begin{equation}}
\newcommand{\eeq}{\end{equation}}
\def\bea{\begin{eqnarray}}
\def\eea{\end{eqnarray}}
\newcommand{\ba}{\begin{array}}
\newcommand{\ea}{\end{array}}

\def\lsim{\stackrel{<}{{}_\sim}}

\begin{document}

\title{\vskip-2.5truecm{\hfill \baselineskip 14pt {{
\small  \\    \hfill FTUV/99-48 \\
\hfill IFIC/99-50\\
\hfill July 1999}}\vskip .9truecm}
 {\bf Supersymmetric Electroweak Baryogenesis  }}

\vspace{5cm}

\author{Nuria Rius and Ver\'onica Sanz
 \\  \  \\
{\it Depto. de F\'\i sica Te\'orica and IFIC, Centro Mixto }\\
{\it Universidad de Valencia-CSIC, Valencia, Spain}\\
}

\date{}
\maketitle
\vfill

\begin{abstract}
\baselineskip 20pt
We calculate the baryon asymmetry generated at the electroweak phase 
transition in the minimal supersymmetric standard model, using a new
method to compute the CP-violating asymmetry in the Higgsino flux
reflected into the unbroken phase. 
The method is based on a Higgs insertion expansion.
We find that the CP asymmetry at leading order is proportional to the 
change in $\tan \beta$ in the bubble wall, 
which is at most of order $10^{-2}$, while at next-to-leading 
order this suppression factor disappears.
This result may enhance the final baryon asymmetry generated 
during the electroweak phase transition for small 
$\Delta \beta \;(< 10^{-3}$).
\end{abstract}
\vfill
\thispagestyle{empty}

\newpage
\pagestyle{plain}
\setcounter{page}{1}

\renewcommand{\thefootnote}{\arabic{footnote}}
\setcounter{footnote}{0}

\section{Introduction}

The baryon to entropy ratio in the observed part of the 
Universe is constrained to be $n_B/s \sim 10^{-10}$ by 
primordial nucleosynthesis measurements \cite{nuc}. 
Sakharov \cite{sak} established more than thirty years ago 
the three basic requirements for obtaining this baryon asymmetry as 
a result of particle interactions in the early universe, namely 
baryon number violation, C and CP violation and  
departure from thermal equilibrium. 
These conditions may be satisfied at weak scale temperatures, if the
electroweak phase transition is first order \cite{krs}.
Electroweak baryogenesis provides an explanation of 
the observed baryon asymmetry of the Universe (BAU) in terms of 
experimentally accessible physics,  
hence much attention has been devoted to the study of this
possibility \cite{ckn}-\cite{r}.

Although the Standard Model contains all the necessary 
ingredients for electroweak baryogenesis, the phase transition is too 
weakly first order to avoid the wash out of the generated baryon 
asymmetry, for the Higgs mass experimentally allowed \cite{pt}. 
Moreover, the CP asymmetry induced by the Kobayashi-Maskawa phase is 
far too small to account for the observed $n_B / s$ ratio 
\cite{belen,hs}.
Therefore, for the baryon asymmetry to be generated at the 
electroweak phase transition, new physics is required at the 
weak scale.

Among the different extensions of the Standard Model, 
low energy supersymmetry is a well motivated possibility, and 
thus several groups have recently studied under which 
conditions electroweak baryogenesis is feasible in the
framework of the Minimal Supersymmetric Standard Model (MSSM).
Regarding the strength of the phase transition, a region in 
the space of supersymmetric parameters has been found 
where the phase transition is strong enough to avoid the 
wash out of the generated baryon asymmetry \cite{quiros}-\cite{cm}.
Such region corresponds to a light Higgs boson and a light 
top squark, within the reach of LEP2 and Tevatron colliders.
As for the baryon asymmetry, in the region of parameter space favored by 
the previous phase transition studies, it is mainly generated by charginos 
and neutralinos, provided they are not much heavier than the 
critical temperature ($T_c \sim 100$ GeV)\cite{hn,ceta}.

The physics of the mechanism which produces the baryon asymmetry is 
essentially agreed upon, namely particles in the plasma interact 
with the bubble wall and due to CP-violation lead to a chiral asymmetry 
of fermions in front of the wall, which in turn biases sphaleron 
processes to produce the BAU.
Several groups have estimated the baryon asymmetry generated at the 
electroweak phase transition in the MSSM: 
in refs.\cite{hn,ceta}, CP-violating source terms were computed 
and then inserted into a set of diffusion 
equations, with a prescription not motivated by first principles.
Subsequently, the correct definition of the CP-violating sources
was self-consistently derived in \cite{r}, using the 
closed time-path formalism to write down a set of quantum
Boltzmann equations describing the local particle densities.
The CP-violating sources were computed using a mass expansion 
\cite{hn} and a Higgs insertion expansion \cite{ceta,r}.  
In \cite{cjk}, the diffusion equations and the source terms were 
derived together within the WKB quasi-particle approximation.

In \cite{hn,ceta,r}, the source terms are proportional to 
$v_2(z) \partial_z v_1(z) - v_1(z) \partial_z v_2(z)$, which is zero 
for $\tan \beta = v_2/v_1$ constant in the bubble wall.
It has been shown \cite{bubbles,cm} that the angle $\beta$ varies at most 
by a few percent over the wall in the MSSM, thus it results in 
an important suppression factor. 
On the other hand, in \cite{hn} it was mentioned that for 
charginos and neutralinos such $\Delta \beta$ dependence could disappear 
at higher orders in the mass expansion used to calculate the 
source terms, and in \cite{aoki}, CP asymmetries of the reflection and 
transmission amplitudes for charginos are computed numerically,  
and they are non-zero for constant $\tan \beta$
\footnote{The authors of \cite{cjk} also find a non-vanishing 
Higgsino source term for constant $\tan \beta$, but they 
consider a different Higgsino density \cite{cosmo}, namely the 
difference of the two helicity states instead of the sum, which is 
the one considered by all the other groups, including us.}.

Motivated by this discrepancy and the possibility that it is a 
consequence of the lowest order approximation, we have performed a
new computation of the baryon asymmetry generated by charginos 
at the electroweak phase transition within the MSSM. 
Although we introduce a different method to compute the CP-violating 
source terms, it is based on a Higgs insertion expansion similar to 
the one employed in \cite{ceta,r}, thus we expect that our conclusion 
will also apply to that approach.

The remainder of the paper is structured as follows. 
In section 2 we describe the method used to compute the CP-violating
asymmetries, and present the results for Higgsinos at leading and 
next-to-leading order in the Higgs insertion expansion.
In section 3 we compute the baryon number induced by the Higgsino current,
and we conclude in section 4.

\section{CP asymmetry for Higgsinos}

Baryogenesis is fueled by CP asymmetries induced by the advancing 
bubble wall. 
Unremovable CP-odd phases appear in the mass matrices due to either

a) CP-violating interactions in the thermal loops that correct the 
dispersion relations of particles propagating in the plasma \cite{fs}.

b) space-time dependence of the scalar vevs inside the bubble wall
(for more than one Higgs field), which induces space-dependent 
CP-violating phases. These phases cannot be rotated away in two 
adjacent points by the same unitary transformation \cite{hn}.

When present, the second mechanism dominates over the first one, 
since in the first mechanism there are suppression factors coming 
from loops. While in the Standard Model only the first mechanism is 
possible, the second one controls the generation of the baryon 
asymmetry in all the extensions of the Standard Model proposed 
in the literature for electroweak baryogenesis. 
Thus, particle mass matrices acquire a non-trivial
space dependence when bubbles of the broken phase nucleate and
expand during a first order electroweak phase transition. 
This provides fast non-equilibrium CP-violating effects inside 
the bubble walls (thick walls) or in front of the bubble walls 
(thin walls), and may generate a baryon asymmetry through the 
anomalous $(B+L)$-violating processes when particles 
diffuse to the exterior of the bubble, in the unbroken 
phase.

In what follows, we will focus on CP-violating effects in the 
Higgsino current, since they make the dominant 
contribution in the region of the MSSM parameter space preferred
by phase transition studies \cite{ceta}.
However the method presented here is completely 
general, and may be applied to any particle whose mass 
depends on the scalar vevs. 
We will perform the calculation  
in the thin wall regime ($L_w \lsim \ell$, where $L_w$ is the bubble wall
width and $\ell$ the mean free path of the particle). 
This is a reasonable approximation for weakly interacting 
particles, such as charginos. 
For bubble walls thinner than the mean free path, 
the incoming fermions interact with the bubble wall like quantum 
mechanical particles scattering from a potential barrier.
CP-violating interactions with the scalar field result in 
different reflection probability for fermions 
of a given chirality and their corresponding antifermions, 
leading to a CP asymmetry in the reflected chiral number flux 
\cite{ckn}.

The chargino mass matrix is, in the basis of Winos and Higgsinos
($\tilde{W}, \tilde{H}$),
\beq
M_{\chi} = 
\left (
\ba{cc}
m & g v_2/\sqrt{2}  \\ 
g v_1/\sqrt{2} & \mu
\ea \right) \ ,
\eeq
with $v_i$ the spatially varying Higgs field vevs and $m,\mu$ the 
soft supersymmetry breaking parameters, which contain one physical 
CP violating phase ($\mu = |\mu| e^{i\phi}$).
In the symmetric phase ($v_i=0$), Winos and Higgsinos are   
the mass eigenstates, while 
in the bubble wall and the broken phase they mix to form 
the mass eigenstates, denoted $\tilde{\chi_i} \ , i=1,2$. 
We treat the wall as planar, and assume it has reached a static 
configuration in the wall rest frame, with the vevs of the scalar 
fields being functions only of the $z$ coordinate. 

The quantity of interest is the CP-violating 
asymmetry in the Higgsino current reflected into the unbroken phase. 
In the rest frame of the bubble wall, it is given by 
\bea
j_{CP} &=& j^{r}_{\tilde{W} \rightarrow \tilde{H}} 
     + j^{r}_{\tilde{H} \rightarrow \tilde{H}} 
 + \sum_{i=1}^2 j^{tr}_{\tilde{\chi_i} \rightarrow \tilde{H}} \\
& =& 
\int \frac{d^3 p}{(2 \pi)^3} \; 
[|R^u_{\tilde{W} \rightarrow \tilde{H}}(-p_m,p_{\mu})|^2 - 
|\bar{R}^u_{\tilde{W} \rightarrow \tilde{H}}(-p_m,p_{\mu})|^2] 
\;\frac{p_m}{E} \rho^u(E,p_m) \nonumber\\
& +&  \int \frac{d^3 p}{(2 \pi)^3} \;
[|R^u_{\tilde{H} \rightarrow \tilde{H}}(-p_{\mu},p_{\mu})|^2 - 
|\bar{R}^u_{\tilde{H} \rightarrow \tilde{H}}(-p_{\mu},p_{\mu})|^2] 
\;\frac{p_{\mu}}{E} \rho^u(E,p_{\mu}) \nonumber\\
&+ &  \int \frac{d^3 p}{(2 \pi)^3} \; \sum_{i=1}^2
[ |T^b_{\tilde{\chi_i} \rightarrow \tilde{H}}(p_i,p_{\mu})|^2
- |\bar{T}^b_{\tilde{\chi_i} \rightarrow \tilde{H}}(p_i,p_{\mu})|^2] \; 
\frac{p_i}{E} \rho^b(E,p_i),
\label{jcp} 
\eea
where
\beq
p_m =\sqrt{E^2-p_\parallel^2 - m^2} \ ,
\hspace{1.cm}
p_{\mu}= \sqrt{E^2-p_\parallel^2 - \mu^2} \ , 
\hspace{1.cm}
p_i=\sqrt{E^2-p_\parallel^2 - m_i^2}
\eeq
are the absolute value of the $z$ component of the Wino, 
Higgsino and chargino momenta, and 
\beq
\rho^u(E,p_z) = \frac{1}{e^{(E+v_w p_z)/T} + 1} \ , 
\hspace{1.5cm}
\rho^b(E,p_z) = \frac{1}{e^{(E-v_w p_z)/T} + 1} \ ,
\label{dist} 
\eeq
are the thermal distributions of the charginos in the unbroken (u) and
broken (b) phases, as seen from the rest frame of the wall.
$v_w$ is the wall velocity, $v_w \sim 0.1$ in the MSSM.
The use of equilibrium particle distributions is a good approximation,  
because any departure from thermal equilibrium is caused by the passage
of the wall, and therefore is ${\cal O}(v_w)$. Since we will see that
the final CP asymmetry is already linear in $v_w$, working with 
thermal equilibrium distribution functions amounts to ignore terms of 
higher order in $v_w$ \cite{hn}.

It is well known that one-loop self-energy corrections to the 
propagator modify the dispersion relations of the particles in the 
plasma. The main thermal effects are that particles propagating in 
the plasma acquire an effective mass (even if they are massless in 
vacuum) and have a finite life-time (damping), due to incoherent thermal 
scattering with the medium.
The damping rate, $\gamma$, is defined as (minus) the imaginary part 
of the solution $\omega = \omega({\bf  k})$ of the dispersion relation.

For the Higgsino and Wino, the effective plasma masses in the thermal 
bath may be well approximated by their value in the present vacuum 
($m^2_{\tilde{H}}(T) \simeq |\mu|^2 \; , m^2_{\tilde{W}}(T) \simeq m^2$).
However, it has been shown that the effects of damping 
can lead to a sizeable suppression of the CP asymmetry \cite{belen,hs}, 
due to the loss of coherence of the wave function.
The damping rate of Winos and Higgsinos has been 
estimated in \cite{damping} to be 
$\gamma_{\tilde{H}} \simeq 0.025 T \; ,
\gamma_{\tilde{W}} \simeq 0.065 T$, hence the mean free path 
$\ell \sim 1/(2\gamma) \sim (10-20)/T$
is comparable to the wall width $L_w \sim (20-30)/T$ \cite{bubbles}, 
and decoherence effects may be relevant.

The effects of damping may be taken into account  
by including the imaginary part of the fermion self-energy in the 
dispersion relation \cite{belen,hs,ceta}. In our case, this 
leads to the approximate dispersion relation 
\beq
[\omega({\bf  k}) + i \gamma]^2 = {\bf k}^2 + m^2  \ .
\label{dr1}  
\eeq
If we choose $\omega$ to be real, the momenta must become complex in 
order to satisfy the dispersion relations, and propagation of 
particles in space is damped. We have taken $\omega$ to be real because
energy is conserved in the scattering off the wall, so the reflection
and transmission probabilities are time independent. 
We could have satisfied the dispersion relations with real momenta and
complex $\omega$, but then the reflection and transmission amplitudes 
would have an exponentially decaying time dependence, which would
require us to study the time and space dependence of 
the particle scattering process. 
To obtain an estimate of the damping effects, from (\ref{dr1})
we approximate 
\beq
k \simeq \pm \left\{ \sqrt{\omega^2 - m^2} + i \gamma 
\frac{\omega}{\sqrt{\omega^2 - m^2}} \right\}
\simeq \pm (\sqrt{\omega^2 - m^2} + i \gamma) \ ,
\label{dr2}
\eeq
where $k = |{\bf  k}|$.

We thus calculate the reflection and transmission amplitudes 
at zero temperature (except for the damping), 
using the LSZ reduction formulae in terms of
the propagator in the presence of the bubble wall \cite{np}: 
\begin{eqnarray}
{\cal A} = \int d^4 x \int d^4 y \;e^{-i q_i x} e^{i q_f y} \;\bar{u}(q_f) 
(i \vec{\partial} - \mu) S(y, x) ( - i \vec{\partial} - m)  u(q_i) \nonumber\\
= (2 \pi)^3 \delta(q^x_f - q^x_i) \;\delta(q^y_f - q^y_i) \;\delta(E_f - E_i)\;
A(q^z_i,q^z_f),
\label{lsz}
\end{eqnarray}
with
\beq
S(y, x) \equiv \langle 0| T[ \Psi(y) \bar{\Psi}(x) ] | 0 \rangle .
\label{prop}
\eeq 
An analogous expression holds for antiparticles. The spinors in formula 
(\ref{lsz}) are on-shell and normalized to unit flux in the $z$ direction, 
i.e.
\beq               
\bar{u}\; \gamma_z \;u = 1.
\label{norm}
\eeq

Momenta in the $x$ and $y$ directions are conserved, because the 
potential created by the bubble wall only depends on the $z$ coordinate.
The transmission and reflection amplitudes are then functions only of 
the momenta in the $z$ direction and can be computed in a simpler way 
by first boosting to a  frame where $q_x , q_y = 0$. 
With the proper normalization chosen for the 
spinors (\ref{norm}), the amplitude in the boosted frame is simply given
by (\ref{lsz}), with the propagator and incoming and outcoming momenta 
substituted by the boosted ones. 
 
The expression of $j_{CP}$ can be further simplified by using CPT 
and unitarity constraints, which imply 
\beq
|R^u_{\tilde{W} \rightarrow \tilde{H}}|^2 + 
|R^u_{\tilde{H} \rightarrow \tilde{H}}|^2 + 
\sum_{i=1}^2 \; |T^b_{\tilde{\chi_i} \rightarrow \tilde{H}}|^2 = 1 \ .
\label{uni}
\eeq
Substituting (\ref{uni}) in eq.(\ref{jcp}), and expanding the 
Fermi distributions for small wall velocities we obtain:
\bea
j_{CP} &=& \frac{v_w}{T}
\int \frac{d^3 p}{(2 \pi)^3} \; \left \{
-(p_{\mu}+p_1) [|R^u_{\tilde{H} \rightarrow \tilde{H}}(-p_{\mu},p_{\mu})|^2 - 
|\bar{R}^u_{\tilde{H} \rightarrow \tilde{H}}(-p_{\mu},p_{\mu})|^2] 
\right. \nonumber\\
& -& \left. (p_m + p_1)
[|R^u_{\tilde{W} \rightarrow \tilde{H}}(-p_m,p_{\mu})|^2 - 
|\bar{R}^u_{\tilde{W} \rightarrow \tilde{H}}(-p_m,p_{\mu})|^2] 
\right. \nonumber\\
&+ & \left. (p_2 - p_1)
[ |T^b_{\tilde{\chi_2} \rightarrow \tilde{H}}(p_2,p_{\mu})|^2
- |\bar{T}^b_{\tilde{\chi_2} \rightarrow \tilde{H}}(p_2,p_{\mu})|^2] 
\right \} 
\frac{p_m}{E} \; \rho(E) \;[1 - \rho(E)]  \ ,
\label{jcp2} 
\eea
where $\rho(E) = 1/(e^{E/T} + 1)$ is the Fermi distribution. 
This result explicitly shows that the out of equilibrium condition 
needed for baryogenesis is due to the expansion of the bubble 
wall through the thermal bath. As anticipated, the CP violating current 
$j_{CP}$ is linear in the wall velocity, $v_w$, and the 
use of the equilibrium distributions in the calculation 
is justified.

In order to compute the amplitudes in eq.(\ref{lsz}) we would need the 
exact propagator in the presence of the wall (\ref{prop}). 
In ref.\cite{hn}, the approach was to perform an
expansion in powers of mass, which is effectively an expansion in 
$M(z)/E$, 
but this approximation is not justified, since the region of interest 
is always $E \sim M$. 
Instead, we perform a Higgs insertion expansion \cite{ceta}, which will be 
a good approximation at least close to the symmetric phase. 
By making a phase redefinition of 
the Higgsino field, we can write the mass matrix as
\beq
M_{\chi}(z) = M_{\chi}^0 + \delta M_{\chi}(z)
\eeq
where
\beq
M_{\chi}^0 =
\left (
\ba{cc}
m &  0  \\ 
0 & |\mu|
\ea \right)
\eeq
and
\beq
\delta M_{\chi}(z) =
\left (
\ba{cc}
0 & e^{-i \phi} u_2(z)  \\ 
u_1(z) & 0
\ea \right) \ ,
\eeq
with $u_i(z) = g v_i(z)/\sqrt{2}$. 
Now, expanding in $\delta M_{\chi}(z)$, the approximate
result for the propagator is 
\beq
S(x_2, x_1)= \int \prod_i d z_i \;S^{(0)}(x_2,z_1) \delta M_{\chi}(z_1) 
S^{(0)}(z_1,z_2) \delta M_{\chi}(z_2) \ldots S^{(0)}(z_i,x_1), 
\eeq
where the integration is done over all $z_i (-\infty, \infty)$
and $S^{(0)}$ stands for the propagators of Higgsino and Wino in 
the symmetric phase, at zero temperature. 
When the approximate dispersion relation (\ref{dr2}) is used, 
$S^{(0)}$ includes also the damping. In the boosted frame 
($p_x = p_y = 0$), for the Wino it is
given by

\bea
S^{(0)}(z_2,z_1) &=& 
i \int \frac{dp_z}{2 \pi} 
\frac{e^{i p_z (z_2-z_1)}}{(\omega + i \gamma_{\tilde{W}})^2 - p_z^2 - m^2}
\left( \omega \gamma^0 - p_z \gamma^z + m \right)
\nonumber \\
&\simeq &\frac 1 2 
\left\{ \theta(z_2 - z_1) e^{(i p_m - \gamma_{\tilde{W}})(z_2-z_1)}
\left( \frac{\omega}{p_m} \; \gamma^0 - \gamma^z + \frac{m}{p_m} \right)
\right. \nonumber \\
&+& \left.
\theta(z_1 - z_2) e^{-(i p_m - \gamma_{\tilde{W}})(z_2-z_1)}
\left( \frac{\omega}{p_m} \; \gamma^0 + \gamma^z + \frac{m}{p_m} \right)
\right\}  \ .
\label{propw}
\eea
The Higgsino propagator has an analogous expression, changing 
$m \rightarrow |\mu|$, $p_m \rightarrow p_{\mu}$ and 
$\gamma_{\tilde{W}} \rightarrow \gamma_{\tilde{H}}$. 
Here, $\omega$ is the energy in the boosted frame,
$\omega = \sqrt{p_m^2 + m^2} = \sqrt{p_{\mu}^2 + |\mu|^2}$
and we have only included the damping rate in the exponential factors,
where the effect is expected to be more important.

\subsection{Leading order}

The leading CP violating contributions arise at second order in the 
Higgs insertion expansion, ${\cal O}(v^2)$,  
as expected from refs.\cite{hn,ceta} where this was also the order of
the CP violating currents.
Since the perturbation matrix $\delta M_{\chi}(z)$ is off-diagonal 
in the Wino-Higgsino basis, at lowest order 
only Winos can be reflected or transmitted into Higgsinos,
and we get  
\beq
R^{(1)}_{\tilde{W} \rightarrow \tilde{H}}(-p_m,p_{\mu}) 
=\bar{u}_{\tilde{H}}(q_f) \int_0^\infty dz \,
e^{i(p_m + p_{\mu}) z} e^{-\gamma z} 
\{u_1(z) L + u_2(z) \, e^{i\phi} R \}
u_{\tilde{W}}(q_i)  \ ,
\label{r1}
\eeq
where $\gamma \equiv \gamma_{\tilde{H}} + \gamma_{\tilde{W}}$, 
$q_i = (\omega, -p_m - i \gamma_{\tilde{W}})$, 
$q_f = (\omega, p_{\mu} + i \gamma_{\tilde{H}})$, 
$\omega$ is the energy in the boosted frame ($q_x=q_y=0$)
and $L,R$ are the chiral projectors. 
The result for the transmission amplitude 
$T^{(1)}_{\tilde{W} \rightarrow \tilde{H}}(p_m,p_{\mu})$
can be obtained just changing $q_i$ by 
$q'_i = (\omega, p_m - i \gamma_{\tilde{W}})$
and $p_m$ by $-p_m$ in eq.(\ref{r1}).

Using standard techniques to compute Dirac matrix traces, and 
substituting the leading order reflection and transmission 
contributions in eq.(\ref{jcp2}), we obtain 
\beq
j_{CP}^{(1)} = \frac{v_w}{T} \; \sin \phi 
\int \frac{d^3 p}{(2 \pi)^3\; E} \; 
F^{(1)}(p_m) \;
\rho(E) \;[1 - \rho(E)] \ ,
\label{jcp3}
\eeq 
where $F^{(1)}(p_m) = - (p_m + p_\mu) F^{(1)}_r(p_m)  + 
(p_m - p_\mu) F^{(1)}_t(p_m)$, being 
\beq
F^{(1)}_r(p_m) = \frac{2 m |\mu|}{p_{\mu}}  
\int dz_1 dz_2 e^{i(p_m+p_{\mu})(z_1-z_2)} e^{-\gamma(z_1-z_2)}
\{ u_1(z_1) u_2(z_2) - u_1(z_2) u_2(z_1) \} 
\label{res1}
\eeq
the contribution coming from the reflection, and 
\beq 
F^{(1)}_t(p_m) = F^{(1)}_r(-p_m) \ .
\label{tr1}
\eeq
the transmission one.
>From the above equations, we see that the CP asymmetry $j_{CP}^{(1)}$
vanishes if 
$\tan \beta = u_2/u_1$ is constant along the bubble wall, in agreement 
with refs.\cite{hn,ceta}, where the computation of the CP violating
currents was done to lowest nontrivial order in the mass and 
Higgs insertion expansions, respectively.

To obtain the final result we must still specify the shape and 
speed of the bubble wall. For the wall profile we take 
the following semirealistic approximation
\footnote{The difference in the final result using the ansatz (\ref{pared}) 
or the functional kinks is typically $5 \%$, although for some values of
the soft masses may be as large as $40 \%$.}
:
\bea
v(z) &=& \frac{v}{2} \left\{1 - \cos \left( \frac{z \pi}{L_w} \right) 
\right\} [\theta(z)-\theta(z-L_w)] + v \; \theta(z-L_w) 
\nonumber \\
\beta(z) &=& \frac{\Delta \beta}{2} 
\left\{1 - \cos \left( \frac{z \pi}{L_w} \right) \right\} 
[\theta(z)-\theta(z-L_w)] + \Delta \beta \; \theta(z-L_w) 
\label{pared}
\eea
where $v^2 = v_1^2 + v_2^2$ and $v_{1,2}$ are the vevs of the two Higgs 
doublets in the broken phase, at the critical temperature. 
For the remaining parameters we take the bubble wall width 
$L_w= 25/T$, the wall velocity $v_w=0.1$ and 
$\Delta \beta = 0.01$ \cite{bubbles,cm}.

Fig. 1 shows $F^{(1)}_t(p_m)$, which is the dominant contribution, 
as a function of $p_m$, for
$v/T_c = 1$ and different values of the Higgsino and Wino masses, 
($|\mu|,m$). The weak coupling at the phase transition
temperature is $\alpha_w = 0.035$.
We can see an enhancement of the CP asymmetry for values of 
$|\mu|$ close to $m$, i.e., when the Higgsino and Wino are nearly 
degenerate. This resonant behavior, also found in \cite{ceta}, 
is enhanced by the approximations made in our calculation, 
because the masses of the particles propagating in the plasma
are those of the symmetric phase, and thus strong degeneracies 
occur when $|\mu| \sim m$. 

\begin{figure} 
\begin{center}
\psfig{figure=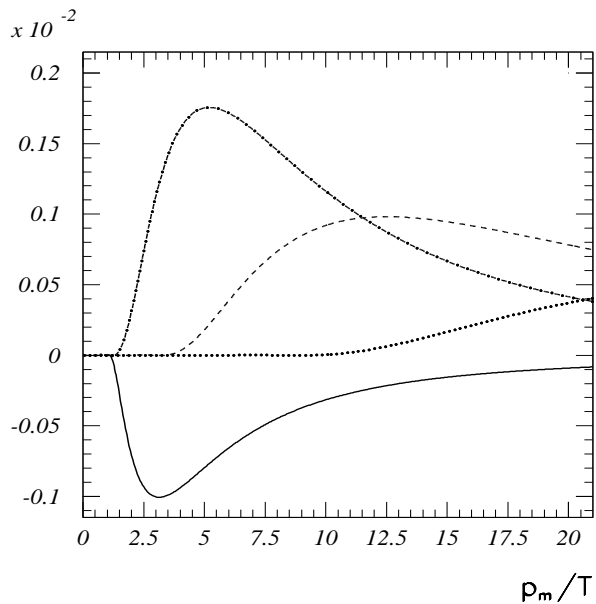,rwidth=2.5in,rheight=2.5in,width=5in,height=6in,bbllx=-30pt,bblly=-30pt,bburx=450pt,bbury=600pt}
\end{center} 
\vspace*{1cm}
\caption{$F^{(1)}_t(p_m)$ as a function of $p_m$ for Higgsino mass 
$|\mu|=1.$
and Wino mass $m = 0.5$ (solid), 
$m = 1.5$ (dashed-dotted), $m = 2.$ (dashed) and 
$m = 3.$ (dotted). 
All in units of the temperature.}
\end{figure}

Regarding the damping effects, as explained before the lifetime 
of the charginos $\ell \sim \gamma^{-1}$ is of the same order of magnitude 
as the wall width, $L_w$.
In this situation, {\it a priori}  it is not clear whether the damping 
will have an important effect or not. 
To estimate it, we have repeated the leading order calculation 
neglecting the damping rate, i.e. setting $\gamma=0$.
The resulting $j_{CP}$ is typically one order of magnitude larger, 
independently of the Higgsino and Wino masses (provided they are
not degenerate~\footnote{For $|\mu| \sim m$, 
since the transmission contribution behaves 
as $1/(p_m - p_\mu)$ the computation 
without damping rate is divergent.}).
Therefore we conclude that although the suppression of the CP 
asymmetry due to the scattering of charginos with 
the plasma is not as enormous as in the case of strongly 
interacting particles (such as quarks in the Standard Model),
decoherence effects are not negligible.

\subsection{Next-to-leading order}

In this section, we will compute the CP violating current $j_{CP}$ 
at next order in the Higgs insertion expansion.
Now both Wino and Higgsino can be reflected or transmitted into 
Higgsino and the CP asymmetry (\ref{jcp2}) 
can be written as
\beq
j_{CP}^{(2)} = \frac{v_w}{T} \; \sin \phi 
\int \frac{d^3 p}{(2 \pi)^3\; E} \; 
F^{(2)}(p_m) \;
\rho(E) \;[1 - \rho(E)]  \ ,
\label{jcp4}
\eeq 
where the function 
$F^{(2)}(p_m)$ is given in Appendix A.

\begin{figure} 
\begin{center}
\psfig{figure=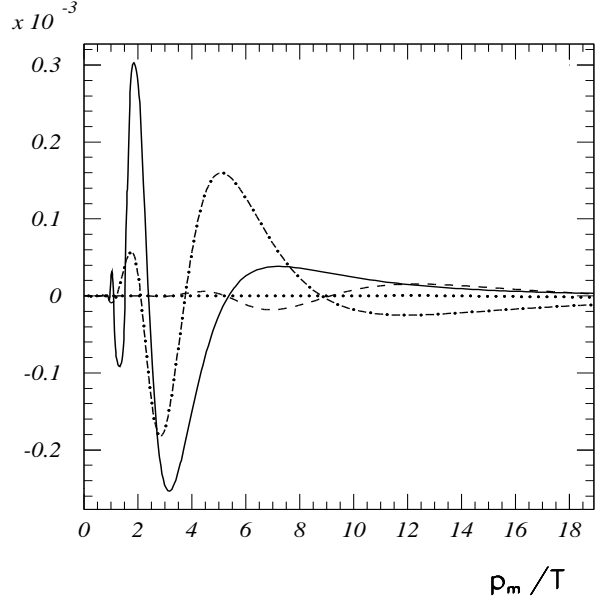,rwidth=2.5in,rheight=2.5in,width=5in,height=6in,bbllx=-30pt,bblly=-30pt,bburx=450pt,bbury=600pt}
\end{center} 
\vspace*{1cm}
\caption{$F^{(2)}_t(p_m)$ as a function of $p_m$ for $\tan \beta = 2$, 
$|\mu|=1.$ and $m = 0.5$ (solid), 
$m = 1.5$ (dashed-dotted), $m = 2.$ (dashed) and 
$m = 3.$ (dotted). 
All masses in units of the temperature.}
\end{figure}

Contrary to the lowest order result, we find that 
$j_{CP}^{(2)} \neq 0$ even if $\tan \beta$ remains constant in the bubble 
wall. Thus the $\Delta \beta$ suppression disappears at higher order
in the Higgs insertion expansion, as mentioned in \cite{hn}, and 
in agreement with the numerical computation of \cite{aoki}.


For constant $v_2/v_1$ we obtain $F^{(2)}(p_m) = 
- (p_m + p_\mu) F^{(2)}_r(p_m) + (p_m - p_\mu) F^{(2)}_t(p_m)$, with
\bea
F^{(2)}_r(p_m) &=&
\frac{m |\mu|}{2 p_{\mu}} \;\sin(4 \beta) \;  
{\rm Im} \left\{  
\int_0^\infty dz_4 u(z_4) e^{-i p_s z_4}  e^{-\gamma z_4} \right.
\nonumber \\
&\times& \left.
\int_0^\infty dz_2 u(z_2) e^{i p_s z_2}  
\int_0^{z_2} dz_1 u(z_1) e^{i p_r z_1} 
\int_0^{z_2} dz_3 u(z_3) e^{-i p_r z_3} \right.
\nonumber \\
&\times& \left.
e^{-\gamma_{\tilde{H}} (z_3+z_2-z_1)} \;
e^{-\gamma_{\tilde{W}} (z_1+z_2-z_3)}
\right \}
\label{res2}
\eea
and
\bea
F^{(2)}_t(p_m) &=& - 
\frac{m |\mu|}{2 p_{\mu}} \; \sin(4 \beta) \;   
{\rm Im} \left\{  
\int_0^\infty dz_4 u(z_4) e^{i p_r z_4} e^{-\gamma z_4} \right. 
\nonumber\\
&\times& \left.
\int_0^\infty dz_3 u(z_3) e^{i p_s z_3}  
\int_0^{z_3} dz_2 u(z_2) e^{-i p_r z_2} 
\int_0^{z_2} dz_1 u(z_1) e^{-i p_s z_1} \right.
\nonumber \\
&\times& \left.
e^{-\gamma_{\tilde{H}} (z_3+z_2-z_1)} \;
e^{-\gamma_{\tilde{W}} (z_1+z_3-z_2)}
\right \} \ ,
\label{tr2}
\eea
where $u(z)= g v(z)/\sqrt{2}$, $p_s = p_m + p_{\mu}$, 
$p_r = p_m - p_{\mu}$ and
$\gamma = \gamma_{\tilde{H}} + \gamma_{\tilde{W}}$.

Again, the dominant contribution comes from the transmission amplitude.
In Fig. 2 we plot $F^{(2)}_t(p_m)$ as a function of $p_m$, 
for the same values and shape of the wall  
profile as in the previous section, except that now 
$\Delta \beta = 0$ and we take $\tan \beta = 2$ (favored by studies 
of the phase transition). Recall that for this choice of 
parameters, the CP asymmetry at lowest order vanishes, so this 
is the leading order result. 
We also find an 
enhancement of the CP asymmetry when $|\mu| \sim m$.

\section{Baryon asymmetry}

Now we have to solve a set of coupled differential 
equations describing the effects of diffusion \cite{diffu}, 
particle number changing 
reactions and CP asymmetries, to obtain the various particle densities 
in the MSSM
\footnote{We neglect the Debye screening of induced gauge charges, 
since the effect on the baryon number produced is ${\cal O}(1)$
\cite{screen}.}.
We follow the approach of refs.\cite{hn,ceta} (see \cite{hn} for 
details). If the system is near equilibrium and the particles interact 
weakly, the particle number densities $n_i$ are given by
\beq
n_i = k_i \mu_i \frac{T^2}{6} \ ,
\eeq
where $\mu_i$ is the local chemical potential for particle species $i$,
and $k_i$ is a statistical factor of order 2 (1) for light bosons 
(fermions), while for particles much heavier that $T$ it is Boltzmann
suppressed.

The particle densities we need to include are the left-handed  
doublet, $q_L \equiv (t_L + b_L)$, the right handed top quark 
$t \equiv t_R$, the Higgs particles 
$h=(H_1^0 + H_1^- + \bar{H}_2^0 + \bar{H}_2^+)$, 
and their superpartners, $\tilde{q}_L, \tilde{t}_R, \tilde{h}$.   
The particle numbers of these species can change due to the 
top quark Yukawa interaction with rate $\Gamma_t$, 
the top quark mass interaction with rate $\Gamma_m$,
the Higgs self-interactions with rate $\Gamma_h$, the strong sphaleron 
interactions with rate $\Gamma_{ss}$, the anomalous weak interactions
with rate $\Gamma_{ws}$ and the gauge interactions (which we shall
assume that are in equilibrium).
Then, the system may be described by the densities 
$Q=q + \tilde{q}$, $T= t + \tilde{t}$ and $H= h + \tilde{h}$. 
Alternatively, one could use the Higgs particle density 
$H' = (H_1^0 + H_1^- + H_2^0 + H_2^+) + $ superpartners, 
which does not have the $\Delta \beta$ suppression in the 
corresponding source term \cite{cjk}.
However, if Yukawa or helicity-flipping interactions were
in equilibrium this would force $H_1 = - H_2$, leading to 
$H'= Q = T = 0$. 
Thus, the approximations we make below would give a vanishing 
baryon asymmetry, and a numerical solution of the 
diffusion equations for the three particle densities including
all the relevant interaction rates is needed
\cite{cosmo}.

CP-violating interactions with the phase boundary produce an 
injected Higgsino flux, which we model as
\beq
J^{inj}(z)= \xi \; j_{CP} \; \delta(z-v_w t)  \ ,
\eeq
where $j_{CP}$ is the net Higgsino flux reflected into the unbroken phase
and $\xi$ defines the persistence length of the current in the vicinity of 
the wall, i.e., it parameterizes our ignorance about how the
injected flux thermalizes. This approximation 
is reasonable if the injected current
thermalizes in a time $\tau_{th}$ short in comparison to the time
the particle spends diffusing before being recaptured by the wall, 
i.e., for small velocities of the wall 
($v_w \ll 1/\sqrt{3}$) \cite{jpt}. 
We use the estimate for $\xi$ of ref.\cite{jpt},  
$\xi \sim 6 D_h \langle v \rangle$, 
where $D_h$ is the diffusion constant of the Higgsino, which can be
approximated by the one of left handed leptons, 
$D_h \sim 110/T$ \cite{jpt}, and $\langle v \rangle$
is the average velocity of the Higgsinos in the reflected flux, 
\beq
\langle v \rangle \equiv \frac{\int \frac{d^3 p}{(2 \pi)^3}
F^{(i)}(p_m) \frac{p_m}{E^2} \rho(E) \;[1 - \rho(E)] }
{\int \frac{d^3 p}{(2 \pi)^3} F^{(i)}(p_m) \frac{1}{E} 
\rho(E) \;[1 - \rho(E)] }  \ .
\eeq
with $F^{(i)}(p_m)$ ($i=1,2$) defined in eqs. (\ref{jcp3}) and 
(\ref{jcp4}), respectively, for the leading and next-to-leading
order computations of the CP asymmetry.

In ref.\cite{r}, the closed time-path formalism was used to derive 
a set of quantum Boltzmann equations describing the local number 
density asymmetries of the particles involved in supersymmetric 
electroweak baryogenesis. In these diffusion equations the 
CP-violating sources which fuel baryogenesis are self-consistently
incorporated.
According to them, the CP-violating source term which should 
be inserted in the diffusion equation for the Higgs density is
given by
\beq
\gamma_{\tilde{h}} \sim \frac{J^{inj}}{\tau} \ ,
\eeq
where $\tau = \gamma_{\tilde{H}}^{-1}$ is the thermalization time of the 
Higgsino.

Assuming that the rates $\Gamma_t$ and $\Gamma_{ss}$ are fast, so that 
$Q/k_Q - H/k_H - T/k_T = {\cal O}(1/\Gamma_{t})$ and 
$2 Q/k_Q - T/k_T + 9(Q+T)/k_B = {\cal O}(1/\Gamma_{ss})$, we obtain 
\bea
Q &=& H \,\,
\frac{k_Q (9 k_T - k_B)}{k_H (9 k_T + k_B + 9 k_Q)} + 
{\cal O}(1/\Gamma_{ss},1/\Gamma_t) \ ,
\nonumber \\
T &=& - H \,\, 
\frac{k_T (9 k_B + 9 k_Q)}{k_H (9 k_T + k_B + 9 k_Q)} + 
{\cal O}(1/\Gamma_{ss},1/\Gamma_t) \ .
\eea
Substituting these expressions 
we find the equation for the Higgs density
\beq
\bar{D} H''  - v_w H' - \bar{\Gamma} H + \bar{\gamma} = 0 \ ,
\label{heq}
\eeq
where $\bar{D}$ is an effective diffusion constant, 
$\bar{\Gamma}$ is an effective decay constant and 
$\bar{\gamma}$ is an effective source term, given by
\footnote {Our expressions slightly differ from those of ref.\cite{hn}, 
but we have checked that this difference is numerically negligible.}  

\bea
\bar{D} &=& \frac{D_q (9 k_Q k_T + k_Q k_B + 4 k_T k_B) + 
D_h (9 k_T + k_B + 9 k_Q) k_H}{9 k_Q k_T + k_Q k_B + 4 k_T k_B + 
 k_H (9 k_T + k_B + 2 k_Q)}  \ , \nonumber \\
\bar{\Gamma}&=&(\Gamma_h + \Gamma_m) \,\, \frac{9 k_T + k_B + 9 k_Q}
{9 k_Q k_T + k_Q k_B + 4 k_T k_B + k_H (9 k_T + k_B + 2 k_Q)}
\ , \\
\bar{\gamma}&=& \gamma_{\tilde{h}} \,\,
\frac{k_H (9 k_T + k_B + 2 k_Q)}
{9 k_Q k_T + k_Q k_B + 4 k_T k_B + k_H (9 k_T + k_B + 2 k_Q)} \ ,
\nonumber
\eea
with $D_q$ ($D_h$) the diffusion constant for quarks 
and squarks (Higgs and Higgsinos).

We are interested in an analytic solution to eq.(\ref{heq}) which
satisfies the boundary conditions $H(\pm \infty) = 0$, and at 
the interphase $z=0$
\beq
H |^+_- = 0   \ ,
\hspace{1.5cm}
\bar{D} H' |^+_- = - \frac{\xi}{\tau} \; \bar{\j}_{CP}  \ ,
\eeq
which are derived by integrating up the eq.(\ref{heq}) through $z=0$, 
imposing the condition that $H$ is at most step-like discontinuous 
across the wall. Here, 
$\bar{\j}_{CP} = (\bar{\gamma}/\gamma_{\tilde{h}}) \;  j_{CP}$.
We use a $z-$ independent effective diffusion constant and a step 
function for the effective decay rate 
$\bar{\Gamma} = \widetilde{\Gamma} \theta(z)$.
The values of $\bar{D}$ and $\bar{\Gamma}$ depend on the 
supersymmetric parameters, for the considered range 
$\bar{D} \sim 0.8 {\rm GeV}^{-1}$,
$\bar{\Gamma} \sim 1.7$ GeV \cite{ceta}.

Then, the solution of eq.(\ref{heq}) in the symmetric phase ($z<0$) 
reads
\beq
H(z) = {\cal A} \, e^{z v_w /\bar{D}} \ ,
\label{sol} 
\eeq
with 
\beq
{\cal A} = \frac{\xi \, \bar{\j}_{CP}}{\tau \lambda_+ \bar{D}}
\label{a}
\eeq
and $\lambda_{\pm} = \frac{v_w \pm \sqrt{v_w^2 + 4 \bar{D} \bar{\Gamma}}}
{2 \bar{D}}$. 

In ref.\cite{hn}, average current densities $J(z)$
were computed in each point $z$ of the bubble wall, and CP-violating 
sources $\gamma_Q(z)$ associated with these currents constructed as 
\beq
\gamma_Q(z) \sim J^0(z)/\tau \ .
\eeq
In order to qualitatively understand the dependence of the
produced baryon asymmetry on the various parameters, the sources 
were approximated as step functions of width $L_w$.
In the thin wall limit ($L_w \rightarrow 0 \ , L_w J^0 =$ constant)
the expression of the coefficient ${\cal A}$ obtained in \cite{hn} 
coincides with ours with the substitution   
$\xi \, j_{CP} \rightarrow L_w J^0$.
This result may help to understand the relation between the two 
approaches, in the regime when both are applicable.

>From the form of (\ref{a}) we see that the CP-violating densities are 
non zero for a time $t \sim {\bar D}/v_w^2$, thus the assumptions 
which lead to the equation (\ref{heq}) for the Higgs density are valid
provided $\Gamma_t, \Gamma_{ss} \gg v_w^2/\bar{D}$.  

What we aim to compute is the total baryon number density left inside 
the bubble, so now we need to consider the effect of the weak sphaleron 
processes, since they provide the only source for net baryon number. 
The equation satisfied by the baryon number density, $n_B$ is 
\beq
D_q n_B'' - v_w n_B' - \theta(-z) n_f \Gamma_{ws} n_L = 0  \ ,
\label{ba}
\eeq
where $n_L$ is the total number density of left handed weak doublet 
fermions, $n_f=3$ is the number of families and for the weak sphaleron 
rate we take $\Gamma_{ws} = 6 \kappa \alpha_w^4 T$ ($\kappa \simeq 1$).
It has recently been estimated as $\Gamma_{ws} \sim C 
\alpha_w \log(1/g) (\alpha_w T)^4$ \cite{bode}, but lattice measurements 
of the rate are consistent with $C \sim 1/\alpha_w$ 
\cite{mt} so this does not affect the numerical value of our 
result.

Assuming that all squarks except $\tilde{t}_{L,R}$ and $\tilde{b}_L$
are heavy, there is no suppression due to the strong sphaleron and 
$n_L$ is given by \cite{hn}
\beq
n_L =  \frac{9 k_Q k_T - 8 k_T k_B - 5 k_Q k_B}
{k_H (9 k_T + k_B + 9 k_Q)} \, H  = \frac{27}{82}  \, H \ .
\eeq
Substituting the Higgs density (\ref{sol}) in eq.(\ref{ba}), we obtain
\beq
\frac{n_B}{s} = - \, \frac {81 {\cal A} \bar{D} \Gamma_{ws}}
{82 v_w^2 s} \ ,
\label{nb}
\eeq
where $s=2 \pi^2 g_* T^3/45$ is the entropy density, with $g_* \sim 126$ 
the effective number of relativistic degrees of freedom. 

In order to compare our approach with previous related work, we 
have calculated the final baryon asymmetry (\ref{nb}) using both,  
the leading and the next-to-leading order results for the 
CP asymmetry in the Higgsino current.
When we use the lowest order CP asymmetry we obtain a baryon number 
to entropy ratio of the same order of magnitude as in \cite{r}
\footnote{We thank A. Riotto for pointing out this fact to us.}, 
where the CP-violating sources were self-consistently 
incorporated in the diffusion equations. 
In previous calculations \cite{hn,ceta}, different definitions of 
the CP-violating currents were used without any first 
principles justification, leading to smaller numerical results
for the final baryon asymmetry
(about one order of magnitude in \cite{hn} and two orders of magnitude 
in \cite{ceta}).

\begin{figure} 
\begin{center}
\psfig{figure=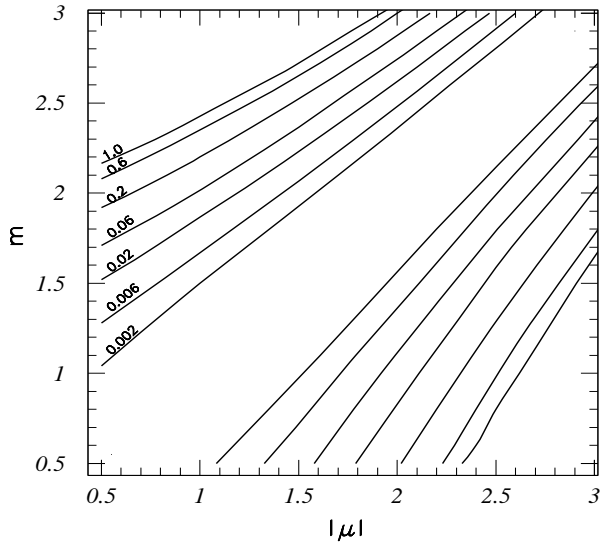,rwidth=2.5in,rheight=2.5in,width=5in,height=6in,bbllx=-30pt,bblly=-30pt,bburx=450pt,bbury=600pt}
\end{center} 
\vspace*{1cm}
\caption{Contour plot of $|\sin \phi \,|$ in the plane $(|\mu|,m)$ for fixed 
$n_B/s \simeq 4 \times 10^{-11}$, according to our leading order 
result for the CP asymmetry in the Higgsino current.
The masses are in units of the temperature.}
\end{figure}

Fig. 3 shows our leading order results.
>From eq.(\ref{nb}), we see that the baryon number produced depends 
linearly on the phase of the Higgsino mass parameter, $\phi$.
So in Fig. 3 we plot the value of $|\sin \phi \,|$ necessary to obtain 
$n_B/s \simeq 4 \times 10^{-11}$, in the $|\mu| - m$ plane, 
using the leading order CP-asymmetry computation
with $\Delta \beta = 0.01$.
The shape of the curves is similar to the ones obtained in
\cite{ceta} but we need a smaller phase to explain the 
observed baryon asymmetry, as discussed above.
We have also calculated the baryon number to entropy ratio 
at leading order for other values of 
the parameter $\Delta \beta$, namely $\Delta \beta = 0.001, 0.005$,
an we find that it scales as 
$n_B/s \propto \Delta \beta$.

For $\tan \beta$ constant in the bubble wall, the CP asymmetry at
lowest order vanishes, and at next-to-leading order 
we obtain 
\beq 
\frac{n_B}{s} \sim (10^{-11} - 10^{-8}\,) \sin \phi \ ,
\eeq
where the largest values correspond to nearly degenerate 
Wino and Higgsino.
In most cases, $|\sin \phi \,|$ should be of order 0.1-1
to explain the observed baryon asymmetry. 
These results correspond to $\tan \beta =2$, but they can easily 
be converted to other values of $\tan \beta$, 
since the $\beta$ dependence is just a global factor 
(see eqs.(\ref{res2}) and (\ref{tr2})).
The values above seem to indicate that 
next-to-leading order corrections are relevant 
only for very small $\Delta \beta (< 0.001$).

\section{Conclusions}

We have computed the baryon asymmetry generated at the electroweak 
phase transition in the MSSM. The leading CP asymmetry which 
fuels baryogenesis is in the Higgsino current, provided that 
Higgsinos and gauginos are not much heavier than the electroweak 
critical temperature ($T_c \sim 100$ GeV). 
The main motivation for the calculation was to settle the question of 
whether or not the $\Delta \beta$ dependence of the CP-violating Higgsino 
current was a consequence of the approximations used in
the computations \cite{hn,ceta,r}. 
Given that $\Delta \beta$ is at most $2.5 \times 10^{-2}$ in the 
MSSM \cite{bubbles,cm}, such dependence leads to a significant
suppression of the produced baryon asymmetry.

We have done the calculation in the thin wall regime, 
using a new method to compute the CP asymmetry by expanding in 
the Higgs vevs. At leading order, our results show the same 
parametric dependence as previous related approaches
\cite{hn,ceta}, in particular the CP asymmetry vanishes if 
$\tan \beta = v_2/v_1$ is constant along the bubble wall. 
However, at next-to-leading order we find a non-zero 
CP-asymmetry even if $\tan \beta$ is constant,  
in agreement with the numerical calculation of  
\cite{aoki}.
Although our approach differs from those of refs.\cite{hn,ceta}, 
all of them contain a Higgs insertion expansion and 
thus we believe that this result will also apply to 
their methods.

We have estimated the effects of damping in the CP asymmetry, 
using a simple model for decoherence.
In the leading order calculation 
we find a suppression of order $0.1$ due to the incoherent scattering 
of charginos with the particles in the plasma. 

The subsequent solution of the diffusion equations 
using the leading order CP asymmetry leads to
a baryon number to entropy ratio in agreement with 
observation, even for small CP-violating phase 
of the $\mu$ parameter, $|\sin \phi| \sim (10^{-2}-10^{-4})$. 
These results are comparable with those of ref.\cite{r}, for the 
same range of the relevant MSSM parameters.  
At this order in the Higgs insertion expansion, 
the dependence of $n_B/s$ with $\Delta \beta$ is approximately linear.

For $\tan \beta$ constant in the bubble wall, the CP asymmetry at
lowest order vanishes, and the next-to-leading order 
calculation gives a baryon asymmetry sufficiently large 
if the CP-violating phase is in the range 
$|\sin \phi| \sim (0.1-1)$. Smaller values are acceptable only 
in the region $m \sim |\mu|$.

The various approximations made in the current analysis lead 
to a sizeable uncertainty in the final result. 
Since the mean free path for charginos $\ell \sim L_w$ 
(bubble wall width), 
we expect the thin wall limit to give an ${\cal O}(1)$ estimate 
of the true solution.
Regarding the Higgs insertion expansion, our next-to-leading
order calculation shows that the leading order result  
is reliable, except for very small values of 
$\Delta \beta (< 0.001$).
We have also made a number of simplifications of the 
diffusion equations \cite{hn}, which are valid 
provided $\Gamma_t, \Gamma_{ss} \gg v_w^2/\bar{D}$, 
$\Gamma_{ws} \ll v_w^2/\bar{D}$ and the scattering 
processes due to Yukawa couplings other than top 
are slow. Given our poor knowledge of  
the parameters involved (diffusion constants $D$,   
reaction rates $\Gamma_{t,ss,ws}$ and wall velocity $v_w$) 
it is not possible to ensure that these inequalities are 
satisfied, but the present estimates seem to indicate so. 

Our most severe approximation was the insertion of the 
CP-violating source in the diffusion equation for the 
Higgs density (\ref{heq}). 
We have modeled the injected Higgsino flux as a delta function 
localized at $z= v_w t$.
The largest uncertainty comes from our ignorance about 
how this injected flux thermalizes. 
Following \cite{jpt}, we have parameterized it by $\xi$, the 
persistence length of the injected current, estimated to be   
$\xi \sim 6  D_h \langle v \rangle$, where 
$6 D_h$ is the velocity randomization time of a diffusing
Higgsino and $\langle v \rangle$ is the average 
velocity of the injected flux. 
To calculate this parameter more precisely involves going 
beyond the diffusion approximation to determine 
exactly how the reflected asymmetry enters in the diffusion 
equation \cite{r}.
Taking into account the above uncertainties and 
approximations, 
we estimate that our results for the baryon asymmetry are 
reliable to about one order of magnitude.

Finally, we want to comment on the differences between our approach 
and the one in refs.\cite{ceta,r}, since both are based on a Higgs 
insertion expansion. 
Our approach is only applicable in the thin wall regime, and we perform 
a tree level computation of the reflection and transmission amplitudes 
to obtain the Higgsino current, much as in \cite{hn}.
On the other hand, the method of \cite{ceta,r} is valid for any thickness
of the bubble wall and they calculate directly CP-violating currents
using the closed-time path formalism, which involves a one-loop 
computation at finite temperature. 
However, at least in the thin wall regime, both approaches should be 
somehow related, and it will be very interesting to understand 
the connection between them.

\section*{Acknowledgments}

We are indebted to P. Hern\'andez for very illuminating discussions. 
We also thank J. Cline, K. Kainulainen, A. Nelson, M. Quir\'os and 
A. Riotto for useful conversations. 
This work was supported in part by DGESIC under Grant No. 
PB97-1261, by DGICYT under contract PB95-1077 and by EEC under the 
TMR contract ERBFMRX-CT96-0090.

\vspace{1cm}

\setcounter{section}{0}
\setcounter{equation}{0}
\renewcommand{\thesection}{Appendix \Alph{section}.}
\renewcommand{\theequation}{A. \arabic{equation}}

\section{}

We present here the different contributions to the 
function $F^{(2)}(p_m)$ 
which appears in the CP asymmetry at next-to-leading order in the Higgs 
insertion expansion, $j_{CP}^{(2)}$ (\ref{jcp4}).
Taking into account the unitarity relation (\ref{uni}), we only need 
the contributions from the reflection and transmission 
of Winos into Higgsinos, and the reflection of Higgsinos into
themselves. 
We work in the boosted frame ($q_x = q_y = 0$). 
In the reflection, the momenta of the initial particles are  
$q_i = (\omega, -p_\mu - i \gamma_{\tilde{H}})$ for Higgsino and 
$q_i = (\omega, -p_m - i \gamma_{\tilde{W}})$ for Wino,  
while $q_f=(\omega, p_\mu + i \gamma_{\tilde{H}})$ 
is the momentum of the Higgsino reflected into the symmetric phase.  

>From $\tilde{H} \rightarrow \tilde{H}$ reflection we obtain:
\bea
F_{\tilde{H}}^{(2)}(p_m) & =& - 2 \; p_\mu \;
F_{r,\tilde{H}}^{(2)}(p_m)
\nonumber \\
& =& -2 i m |\mu| \;
\int dz_1 \ldots dz_4 \int \frac{dp_z}{2 \pi} \frac{dp'_z}{2 \pi}
\frac{e^{(i p_\mu -\gamma_{\tilde{H}})(z_1+z_2)}
e^{(-i p_\mu -\gamma_{\tilde{H}})(z_3+z_4)}
e^{-i p_z (z_2-z_1) + i p'_z (z_4-z_3)}}
{(p^2 - m^2 + i \gamma_{\tilde{W}}) (p'^2 - m^2 - i \gamma_{\tilde{W}})}
\nonumber \\
&\times &\{ 2 m |\mu| \cos \phi [u_2(z_1) u_1(z_2) u_2(z_3) u_1(z_4) - 
u_1(z_1) u_2(z_2) u_1(z_3) u_2(z_4)] \nonumber \\
& & +(q_f \cdot p') [u_2(z_1) u_1(z_2) u_2(z_3) u_2(z_4) -
u_1(z_1) u_2(z_2) u_1(z_3) u_1(z_4) ] \nonumber \\
& & +(q_i \cdot p) [u_2(z_1) u_2(z_2) u_2(z_3) u_1(z_4) -
u_1(z_1) u_1(z_2) u_1(z_3) u_2(z_4) ] \nonumber \\
& & +(q_f \cdot p) [u_1(z_1) u_1(z_2) u_2(z_3) u_1(z_4) -
u_2(z_1) u_2(z_2) u_1(z_3) u_2(z_4)] \nonumber \\
& & +(q_i \cdot p') [u_2(z_1) u_1(z_2) u_1(z_3) u_1(z_4) -
u_1(z_1) u_2(z_2) u_2(z_3) u_2(z_4)] \}  \ ,
\eea
where $p^2 = \omega^2 - p_z^2$, $p'^2 = \omega^2 - p'^2_z$.
By using symmetry arguments, it is easy to see that the above expression 
vanishes identically. 

The contribution coming from 
$\tilde{W} \rightarrow \tilde{H}$ transitions can be written as
\beq
F_{\tilde{W}}^{(2)}(p_m) = - (p_m + p_\mu) F_{r,\tilde{W}}^{(2)}(p_m)
+ (p_m - p_\mu) F_{t,\tilde{W}}^{(2)}(p_m) \ ,
\eeq 
where 
\bea
F_{r,\tilde{W}}^{(2)}(p_m) & =&
\nonumber \\
& =& \frac{2m |\mu|}{p_\mu} \; {\rm Im} \bigg (
\int dz_1 \ldots dz_4 \int \frac{dp_z}{2 \pi} \frac{dp'_z}{2 \pi}
\frac{e^{(i p_\mu -\gamma_{\tilde{H}})z_3 + (i p_m -\gamma_{\tilde{W}})z_1}
e^{[-i(p_\mu + p_m)-\gamma] z_4}
e^{i p_z (z_3-z_2) + i p'_z (z_2-z_1)}}
{(p^2 - m^2 + i \gamma_{\tilde{W}}) (p'^2 - \mu^2 + i \gamma_{\tilde{H}})}
 \nonumber \\
&\times & 
\{ 2 m |\mu| \cos \phi [u_2(z_1) u_1(z_2) u_2(z_3) u_1(z_4) - 
u_1(z_1) u_2(z_2) u_1(z_3) u_2(z_4)]  \nonumber \\
& &+  
(q_f \cdot q_i) [u_2(z_1) u_1(z_2) u_2(z_3) u_2(z_4) -
u_1(z_1) u_2(z_2) u_1(z_3) u_1(z_4) ]  \nonumber \\
& &+  
(q_f \cdot p)[u_2(z_1) u_1(z_2) u_1(z_3) u_1(z_4) -
u_1(z_1) u_2(z_2) u_2(z_3) u_2(z_4) ]  \nonumber \\
& &+  
(q_i \cdot p') [u_1(z_1) u_1(z_2) u_2(z_3) u_1(z_4) -
u_2(z_1) u_2(z_2) u_1(z_3) u_2(z_4)] \nonumber \\
& &+  
(p \cdot p') [u_2(z_1) u_2(z_2) u_2(z_3) u_1(z_4) -
u_1(z_1) u_1(z_2) u_1(z_3) u_2(z_4)] \} \bigg ) \ ,
\label{r2}
\eea
with $\gamma = \gamma_{\tilde{W}} + \gamma_{\tilde{H}}$.  
The contribution of the transmission from the broken phase, 
$F_{t,\tilde{W}}^{(2)}(p_m)$,
can be obtained from eq.(\ref{r2}) changing 
$q_i \rightarrow q'_i = (\omega, p_m - i \gamma_{\tilde{W}})$ and 
$p_m \rightarrow -p_m$ in the exponentials. 
We have only included the damping rate in the exponential factors.

\end{document}